\documentstyle[sprocl,epsfig]{article}

\def\Journal#1#2#3#4{{#1} {\bf #2}, #3 (#4)}

\def\NPB{{\em Nucl. Phys.} B}

\def\PRL{\em Phys. Rev. Lett.}
\def\PRD{{\em Phys. Rev.} D}

\begin{document}

\title{ THERMODYNAMICS OF LATTICE QCD WITH 2 QUARK FLAVOURS:
        CHIRAL SYMMETRY AND TOPOLOGY.
        \footnote{talk presented by D.~K.~Sinclair at Workshop on 
        Nonperturbative Methods in Quantum Field Theory, 2nd -- 13th
        February, 1998, Adelaide, South Australia.}}

\vspace{1in}

\author{J.-F.~LAGA\"{E}, D.~K.~SINCLAIR}

\address{HEP Division, Argonne National Laboratory, 9700 South Cass Avenue,
         Argonne, IL 60439, USA}

\author{J.~B.~KOGUT}

\address{Physics Department, University of Illinois, 1110 West Green Street,
         Urbana, IL 61801, USA}

\maketitle\abstracts{We have studied the restoration of chiral symmetry in
lattice QCD at the finite temperature transition from hadronic matter to a
quark-gluon plasma. By measuring the screening masses of flavour singlet and
non-singlet meson excitations, we have seen evidence that, although flavour
chiral symmetry is restored at this transition, flavour singlet ($U(1)$) axial
symmetry is not. We conclude that this indicates that instantons continue to
play an important r\^{o}le in the quark-gluon plasma phase.}

\section{Introduction}

QCD thermodynamics describes hadronic/nuclear matter at finite temperature
and/or baryon number density. This is relevant to the early universe, neutron
stars and relativistic heavy ion collisions (RHIC).

Hadronic matter undergoes a transition to a quark-gluon plasma at a temperature
$T = T_c \approx 100$--$150$~MeV \cite{milc,htmcgc}. This transition is
believed to be a second order phase transition when $m_u,m_d \rightarrow 0$
($m_s$ appears to be too large to make this transition first order). From now
on we shall ignore the $s$ quark.

At this phase transition the $SU(2) \times SU(2)$ chiral flavour symmetry is
restored. This means that in the plasma phase the effective (screening) masses 
of the $\pi$ and $\sigma(f_0)$ become equal as do the masses of the 
$\eta/\eta'$ and $\delta(a_0)$. The question remains as to whether the anomalous
$U(1)_{axial}$ symmetry is also restored at this transition \cite{pw,shuryak}
If so, all 4 screening masses would become equal above the transition. The
rest of this talk presents lattice measurements aimed at resolving this
question, and discussion of how this is related to topological charge and
fermion zero modes.

If the $U(1)_{axial}$ symmetry is not restored at the chiral(deconfinement)
transition, this means that instantons still play an important r\^{o}le in
the plasma phase. 

The functional integral for a field theory at Euclidean time in a region which
is infinite in each of the spatial directions but has a finite extent, $1/T$,
in the time direction with periodic (antiperiodic) boundary conditions, is the
canonical partition function for the field theory at temperature $T$. This
allows us to simulate lattice QCD at finite $T$ by simulating it on a lattice
whose spatial dimensions are much larger than its time dimension.

This talk presents work which is reported in detail in \cite{kls},
which should be consulted for a more complete description and for more complete
references. For related work which draws similar conclusions see 
\cite{bielefeld,milc2}.

\section{Meson propagators and Topological Charge at finite Temperature}

In QCD, mesons can be created or annihilated by operators quadratic in the
quark fields. Thus a meson propagator has the form
\begin{equation}
     G(x,y) = \langle\bar{\psi}(x){\cal O}^{\dag}\psi(x) 
              \bar{\psi}(y){\cal O}\psi(y)\rangle           .
\end{equation}
For the $\pi$, ${\cal O}=\gamma_5{\bf \tau}$, for the $\sigma$, ${\cal O}=1$,
for the $\eta/\eta'$, ${\cal O}=\gamma_5$ while for the $\delta$, 
${\cal O}={\bf \tau}$.

\begin{figure}[htb]
\epsfysize=4in
\centerline{\epsffile{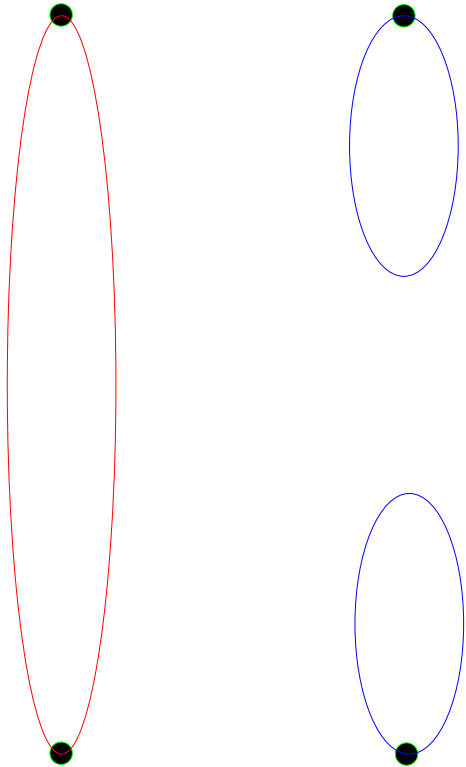}}
\caption{Connected (left) and disconnected (right) contributions to
         meson propagators in a fixed background gauge-field (not shown).
         \label{fig:propagators}}
\end{figure}

The connected parts of the flavour singlet and corresponding non-singlet meson
propagators are the same. In addition to the connected part, the flavour singlet
meson propagator has a disconnected part. These are shown schematically in
figure~\ref{fig:propagators} where the gluon fields, containing the effects of
closed quark loops have been omitted. If this disconnected contribution to the
$\sigma$ and $\eta/\eta'$ propagators were to vanish in the plasma phase, then
the $U(1)_{axial}$ symmetry would be restored.

When the $SU(2) \times SU(2)$ flavour symmetry is restored,
\begin{equation}
           G_\pi = - G_\sigma
\end{equation}
and
\begin{equation}
           G_{\eta'} = - G_\delta
\end{equation}
If the $U(1)_{axial}$ symmetry were restored, 
\begin{equation}
           G_\pi = - G_\delta
\end{equation}
and
\begin{equation}
           G_\sigma = - G_{\eta'}
\end{equation}
in addition.

Now let us discuss the r\^{o}le of instantons. These arguments are appropriate
to the high temperature, quark-gluon plasma phase, where we can take the
fermion mass to zero at finite volume, and we expect a gap between zero modes
and non-zero modes at mass=0. The spectral decomposition of the quark
propagator is
\begin{equation}
         S(x,y) = \sum_\lambda {\psi_\lambda(x) \psi^\dag_\lambda(y) \over
                                                i\lambda + m}
\end{equation}
Now let us return to the consideration of meson propagators. For configurations
with no zero modes ($\lambda \ne 0$), only the connected contributions survive
the chiral ($m \rightarrow 0$) limit. For configurations with one zero mode,
\begin{equation}
         S(x,y) \sim 1/m
\end{equation}
and the dominant contributions to both the connected and disconnected terms
are $\sim 1/m^2$. Such configurations contribute with weight $\propto m^2$
(from the fermion determinant), and so we get a finite contribution. The
fermion determinant suppresses contributions from configurations with $> 1$
zero modes. We find $G_\pi = - G_\sigma$ and $G_{\eta'} = - G_\delta$, i.e.
chiral $SU(2) \times SU(2)$ flavour symmetry is restored.

The relationship to topology is expressed by the Atiyah-Singer index theorem
\cite{as}:
\begin{equation}
         m \int d^4 x {\rm Tr} \gamma_5 S = n_L -n_R = Q_{top}
         = {1 \over 32\pi^2} \int d^4 x {\rm Tr} F \tilde{F}, 
\end{equation}
for a given gauge field configuration. This tells us that the topological
charge is the difference between the number of left-handed and the number of
right-handed zero modes. 

A corollary is
\begin{equation}
         \chi_{top} = {\langle Q_{top}^2 \rangle \over V}
                    = m^2 {\int d^4 x \int d^4 y G_{\eta'}^{dis}(x,y) \over V}
                    = m^2 \chi_5^{dis} 
\end{equation}
where a non-zero value of $\chi_5^{dis}$ would indicate that the $\eta/\eta'$
propagator differed from that of the $\pi$, and that the $U(1)_{axial}$ symmetry
was broken.

\section{Meson propagators on the lattice.}

In lattice QCD, the $SU(2) \times SU(2)$ and $U(1)_{axial}$ symmetries are
explicitly broken. We have worked with staggered fermions where 1 generator
of $SU(2)_{axial}$ remains unbroken. There are no exact zero modes, and one
can only take $m \rightarrow 0$ after taking the continuum limit. We must
therefore work at small but finite $m$.

We work with configurations generated in simulations on a $16^3 \times 8$
lattice in lattice QCD with 2 light ($m/T=0.05$) staggered quark flavours.
Most of these were provided by the HTMCGC collaboration \cite{htmcgc}. The 
chiral transition occurs at $5.475 < \beta_c \equiv 6/g_c^2 < 5.4875$. 

The meson screening masses have been obtained from exponential fits to
propagators for spatial separations. Local operators (noisy point source) were
used for the $\pi$, $\sigma$ and $\delta$. Covariant 4-link operators were
used for the $\pi_4$ and the $\eta'$. $U(1)$ noisy estimators were used to
extract the disconnected contributions to the $\sigma$ and $\eta'$ propagators
\cite{kilcup}.

\begin{figure}[htb]
\epsfysize=4in
\centerline{\epsffile{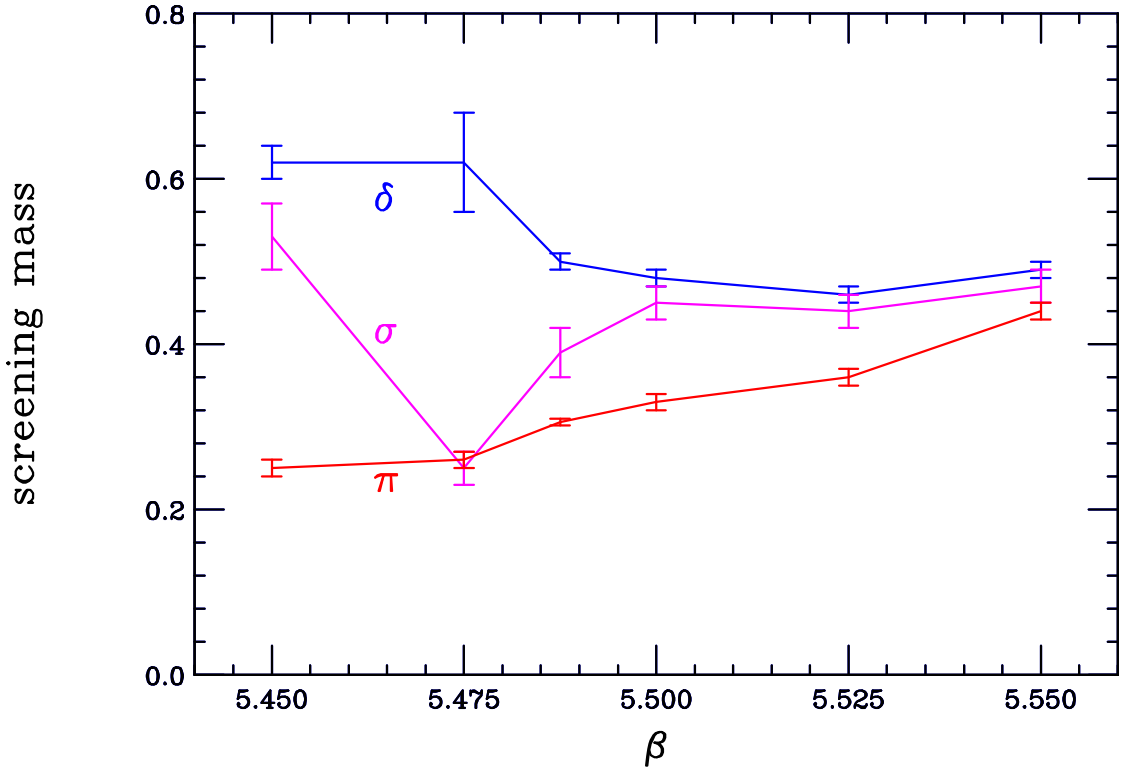}}
\caption{$\sigma$, $\pi$ and $\delta$ screening masses as functions of $\beta$
         for full 2-flavour lattice QCD.\label{fig:masses}}
\end{figure}

Figure~\ref{fig:masses} shows the meson screening masses obtained from
exponential fits to these propagators. (n.b. the flavour non-singlet screening
masses, i.e. those from the connected propagators have been reported in
\cite{htmcgc}.) 
We observe that $m_\sigma$ and $m_\delta$ remain distinct until appreciably
above the transition. Since the $\sigma$ and $\delta$ propagators differ only
by their disconnected pieces, it is these disconnected pieces which are
responsible for this difference. The $\sigma$ and $\pi$ masses are only
distinct above the transition because of the explicit chiral symmetry breaking
from using a finite quark mass.

The scalar disconnected susceptibility (which would be the same as the
pseudoscalar in the symmetric phase in the chiral limit, if we had continuum
symmetries), shows a clear peak at the transition, and is finite well into the
plasma phase.

\section{The fate of zero modes on the lattice.\protect\footnote{The
discussion at the beginning of this section is condensed from Smit and Vink
\protect\cite{sv}}}

Staggered quarks preserve one $U(1)$ subgroup of the $SU(4)_{axial}$ symmetry.
The rest of the $SU(4) \times SU(4)$ chiral symmetry is broken by terms 
${\cal O}(a^2)$. As in the continuum, the eigenvalues of the Dirac operator
have the form:
\begin{equation}
                     i\lambda + m
\end{equation}
and occur in complex conjugate pairs. ``$\gamma_5$'' is an operator which
involves a displacement by 4 links. It does not anticommute with the $m=0$ 
Dirac operator, and does not even have eigenvalues $\pm 1$.

In the continuum,
\begin{equation}
        \langle\lambda|\gamma_5|\lambda\rangle = 0, \;\;\;\; \lambda \ne 0;
\end{equation}
and
\begin{equation}
        \langle 0|\gamma_5|0 \rangle = \pm 1,  \;\;\;\; \lambda = 0.
\end{equation}
For staggered quarks this is no longer true. This is fortunate since all modes
have $\lambda \ne 0$, except on a set of configurations of measure zero. In
order to have a sensible continuum limit, those eigenmodes with $\lambda$
close to zero should have $\langle\lambda|\gamma_5|\lambda\rangle$ close to
a constant which can be renormalized to $\pm 1$, while other eigenmodes should
have $\langle\lambda|\gamma_5|\lambda\rangle$ close to zero, at sufficiently
small coupling. These near-zero eigenmodes should obey
\begin{equation}
                            n_+ - n_- = Q_{top}
\end{equation}
where $Q_{top}$ is an estimate of the topological charge of the configuration,
determined by other means.

\begin{figure}[htb]
\epsfysize=4in
\centerline{\epsffile{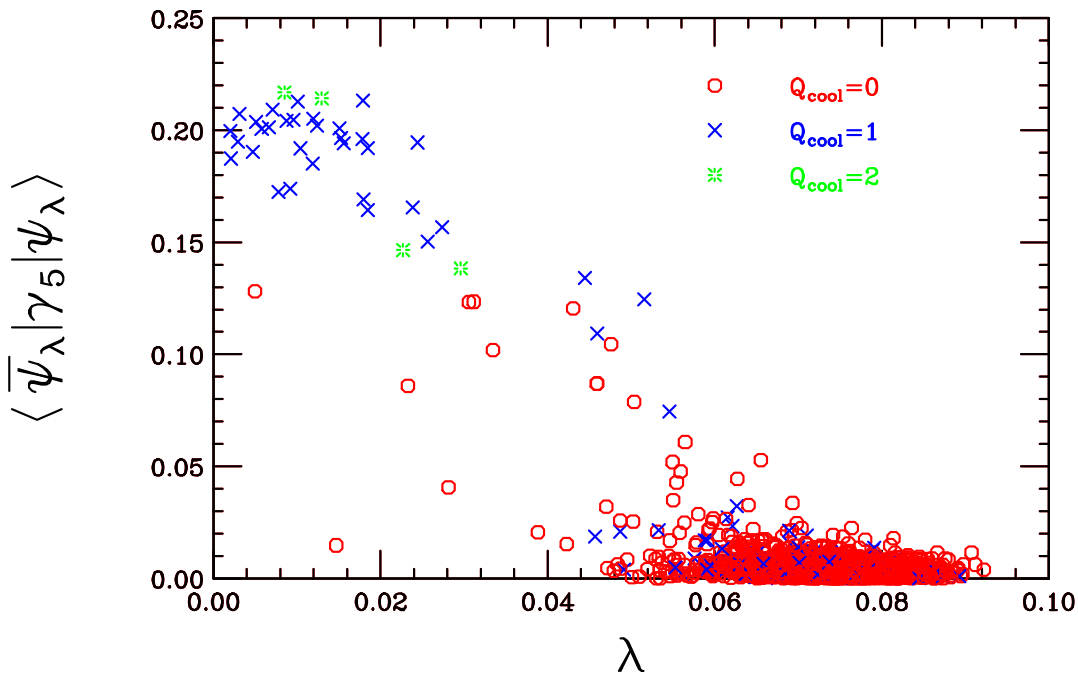}}
\caption{Chirality versus $\lambda$ at $\beta=6.2$(quenched QCD).
         \label{fig:quenched}}
\end{figure}

\begin{figure}[htb]
\epsfysize=4in
\centerline{\epsffile{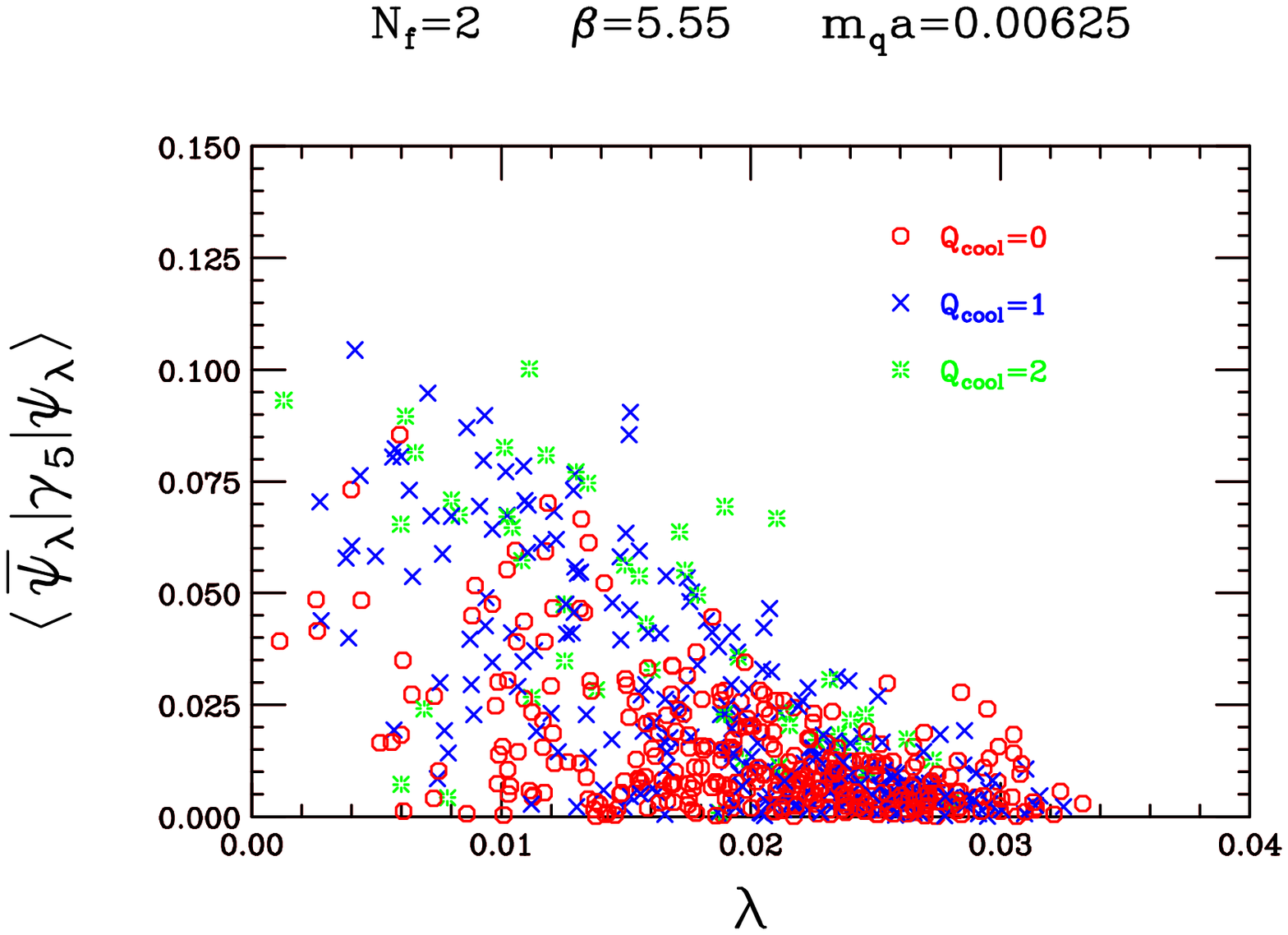}}
\caption{Chirality versus $\lambda$ at $\beta=5.55$(full QCD).
         \label{fig:full}}
\end{figure}

For each of our configurations we have determined the 16 eigenvalues $\lambda$
which are smallest in magnitude, and their corresponding eigenvectors. We have 
repeated this determination for a set of quenched configurations on the same 
size lattice. $\langle\lambda|\gamma_5|\lambda\rangle$ was calculated for
each mode. $Q_{top}$ was measured for each configuration by the cooling method.
We have plotted $\langle\lambda|\gamma_5|\lambda\rangle$ versus $\lambda$ for 
the 8 smallest positive eigenvalues, $\lambda$ for each configuration for a
representative $\beta$ value for our quenched and unquenched simulations in
figures~\ref{fig:quenched},~\ref{fig:full}.

For the quenched case, at $\beta \equiv 6/g^2 = 6.2$ in the high-temperature
phase, it is clear that there is a correlation between small eigenvalues and
large $\langle\lambda|\gamma_5|\lambda\rangle$. In addition, we see that such
eigenvalues tend to occur for configurations with non-zero topological charge.
In the full 2-flavour case at beta=5.55, also in the quark-gluon plasma phase,
there is an indication of similar behaviour. However, since $\beta=5.55$
represents a somewhat larger lattice spacing the effects are far less 
obvious.

\begin{figure}[htb]
\epsfxsize=4in
\centerline{\epsffile{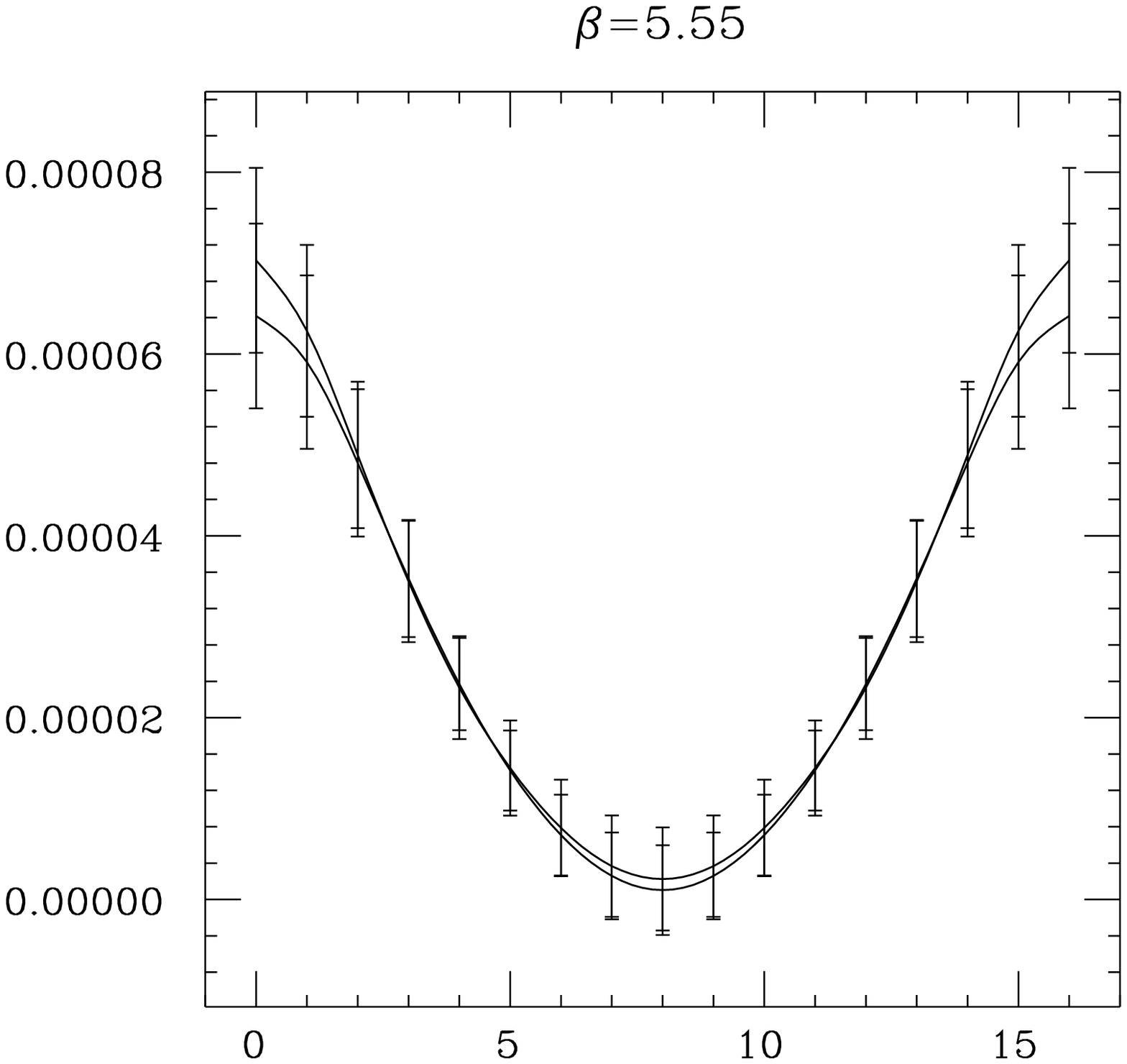}}
\caption{Comparison of the disconnected pseudoscalar correlators 
obtained from a noisy estimator (top curve) and from the truncated 
spectral decomposition of the quark propagator (bottom curve) 
at $\beta=5.55$ ($N_f=2$,$m=0.00625$).\label{fig:ps5p55}}
\end{figure}

We have used these low lying eigenmodes to approximate the quark propagator.
Well above the phase transition this approximation gives a very good 
approximation to the disconnected part of the flavour singlet meson propagators
(see figure~\ref{fig:ps5p55}).
This supports the conclusion that the disconnected part of these propagators
is given entirely by instantons, in the chiral limit.

Because of the finiteness of our would be zero modes, we are forced to remain
at finite quark mass. If we continued to zero quark mass 
$ m \int d^4 x {\rm Tr} \gamma_5 S $ would vanish, as would the disconnected
parts of our singlet meson propagators, leading us to the erroneous conclusion
that the $U(1)_{axial}$ symmetry was restored.

\section{Better Actions}

To get more definitive results would require either using much larger lattices,
or using improved actions which better approximate the flavour chiral symmetries
of the continuum.

As a first try we have used the ``link + staples'' action suggested by the
MILC collaboration \cite{milc3} as a way of improving the flavour symmetry of
the staggered quark action. This improved the approximation to 4-fold
degeneracy of the eigenvalues, which would be present in the continuum.
However, the overall improvement in the
$\langle\lambda|\gamma_5|\lambda\rangle$ versus $\lambda$ plot was not great.

We are now investigating how Wilson, and Wilson + Clover Leaf actions perform
in this regard on the configurations we have. These actions also have chiral
symmetry breaking, but they have exact $SU(2)$ flavour symmetry. The way which
they fail to obey the Atiyah-Singer index theorem is different from the way
staggered fermions fail.

Finally, we are investigating how well domain-wall fermions satisfy the
Atiyah-Singer index theorem and looking at their zero modes on these 
configurations. Domain-wall fermions are 5-dimensional Wilson fermions with
boundary conditions in the 5th dimension which cause light quark states to
reside close to the boundaries in the 5-th dimension \cite{fs}. Their
advantage is that in the limit of infinite 5-th dimension, chiral $SU(2)
\times SU(2)$ flavour symmetry becomes exact, and the approach to this chiral
symmetry appears to be exponential in $N_5$. In this limit, the index theorem
should become exact. Our preliminary calculations on a few configurations show
promise.

Other improved actions, including ones with gauge improvement should be tried.

\section{ Discussion and Conclusions}

At the phase transition from hadronic matter to a quark-gluon plasma in QCD
with 2 zero mass quarks, the chiral $SU(2) \times SU(2)$ chiral flavour
symmetry is restored. We have found evidence that the anomalous $U(1)_{axial}$
symmetry is not restored, but remains broken in the plasma phase. The explicit
breaking of chiral symmetry by the staggered lattice fermions is sufficiently 
large at these lattice spacings to somewhat obscure these results, as does the 
explicit chiral symmetry breaking due to our use of finite quark masses. We
are prevented from decreasing the quark masses to zero by this chiral symmetry
breaking which renders would-be zero modes non-zero. Since our analytic
arguments required a fixed spatial volume, we need to repeat our simulations
on several spatial volumes to study finite size effects, to see that our
conclusions survive in the infinite volume limit.

One can improve on our results by setting the $4 Q_{top}$ eigenvalues of
smallest magnitude to zero by hand. This gives us a non-zero
$ m \int d^4 x {\rm Tr} \gamma_5 S $ and non-zero disconnected contributions to
our singlet meson propagators. However, the chiral flavour symmetry breaking
associated with staggered fermions manifests itself in ways other than the
zero mode shift, and these are not addressed by this procedure. To address
these requires us to study more systematic improvements to finite temperature
simulations and measurements involving light quarks.

If the $U(1)_{axial}$ symmetry remains broken in the high temperature phase,
this means that instantons continue to play an important r\^{o}le in this
phase. In particular they contribute the disconnected part of the propagator
which is responsible for the mass difference between isosinglet and isotriplet
scalar and pseudoscalar mesonic excitations of the plasma. In addition, they
make contributions to the connected meson propagators. This occurs because,
despite the fact that the fermion determinant vanishes as $m^2$ in the chiral
limit for the $Q_{top} = \pm 1$ sector, the connected and disconnected
contributions to the meson propagators for gauge configurations in this sector
diverge as $1/m^2$ giving a net finite contribution. Thus, while 
$m_\pi = m_\sigma$ and $m_{\eta'} = m_\delta$, $m_\pi \ne m_{\eta'}$ and
$m_\sigma \ne m_\delta$. It will be interesting to see if the meson
propagators in the $Q_{top}=0$ sector yield screening masses close to 
$2 \pi T$, the screening mass for a free quark-antiquark pair. If so, it would
indicate that it is the instantons which prevent the quark-gluon plasma from
being described as a gas of free quarks and gluons.

It is important that the chiral flavour symmetry is $SU(2) \times SU(2)$. 
For $SU(3) \times SU(3)$ chiral flavour symmetry, instantons are further
suppressed in the plasma phase, and the $U(1)_{axial}$ symmetry is restored
at the transition, which now becomes first order.

At high temperatures, the disconnected contributions to the flavour singlet
meson propagators are well approximated by the contributions of a few low
lying modes of the quark propagator.

We need improved actions --- especially fermion actions --- to get more
definitive results, unless we have the computing power to work with the much
larger lattices needed to enable us to decrease the lattice spacing enough to
significantly reduce the lattice breaking of chiral flavour symmetry.
Domain-wall fermions appear to be a good bet. It is of interest to know if
such improvements also give one better critical indices at the phase
transition. Others have suggested that the ``perfect action'' improvements are
the most promising \cite{hln}.

Even with the ultimate improved action, currently used simulation methods 
will fair poorly when used to calculate meson propagators in the chiral limit
of the plasma phase. This is because, as mentioned above, configurations which
occur with probability only ${\cal O}(m^2)$, each contribute a term
${\cal O}(1/m^2)$ to the meson propagators. Hence, in the chiral limit,
important, and in fact infinite contributions come from configurations which
never occur. We are studying algorithm modifications to better accommodate
this limit.

\section*{Acknowledgments}
These computations were performed on the CRAY C-90 at NERSC. This work was
supported by the U.~S.~Department of Energy under contract W-31-109-ENG-38,
and the National Science Foundation under grant NSF-PHY96-05199. We thank the
HTMCGC and in particular Urs Heller for the use of their configurations.

\end{document}